\documentclass[12pt]{article} 

\usepackage{epsfig}
\usepackage{graphicx}
\usepackage{comment}
\usepackage{latexsym}
\usepackage{hyperref}
\usepackage{amsmath}
\usepackage{color}
\usepackage{amsbsy}
\usepackage{amssymb}
\usepackage{amsthm}
\usepackage{amsfonts}
\usepackage{cite}

\newcommand{\R}{\mathbb{R}}
\newcommand{\Z}{\mathbb{Z}}

\renewcommand{\O}{{\cal O}}

\newcommand{\ie}{{\em i.e.} }

\newcommand{\where}{\mbox{where}}

\renewcommand{\and}{\mbox{and}}

\newcommand{\espD}{\phantom{\!\!\underset{\displaystyle |}{\cdot}}}
\newcommand{\espDD}{\phantom{\!\!\underset{\displaystyle |}{|}}}

\newcommand{\F}{{\cal F}}
\newcommand{\N}{{\cal N}}

\renewcommand{\P}{{\cal P}}

\newcommand{\M}{{\cal M}}

\newcommand{\Ms}{M_{\rm s}}

\newcommand{\nF}{n_{\rm F}}
\newcommand{\nB}{n_{\rm B}}
\newcommand{\V}{{\cal V}}
\newcommand{\Vone}{\V_{\text{1-loop}}}

\newcommand{\be}{\begin{equation}}
\newcommand{\ee}{\end{equation}}

\newcommand{\bm}{\boldmath} 

\catcode`\@=11
\def\marginnote#1{}
\newcount\hour
\newcount\minute
\newtoks\amorpm
\hour=\time\divide\hour by60 \minute=\time{\multiply\hour by60
\global\advance\minute by-\hour}
\edef\standardtime{{\ifnum\hour<12 \global\amorpm={am}%
        \else\global\amorpm={pm}\advance\hour by-12 \fi
        \ifnum\hour=0 \hour=12 \fi
        \number\hour:\ifnum\minute<10 0\fi\number\minute\the\amorpm}}
\edef\militarytime{\number\hour:\ifnum\minute<10 0\fi\number\minute}
\def\draftlabel#1{{\@bsphack\if@filesw {\let\thepage\relax
   \xdef\@gtempa{\write\@auxout{\string
      \newlabel{#1}{{\@currentlabel}{\thepage}}}}}\@gtempa
   \if@nobreak \ifvmode\nobreak\fi\fi\fi\@esphack}
        \gdef\@eqnlabel{#1}}
\def\@eqnlabel{}
\def\@vacuum{}
\def\draftmarginnote#1{\marginpar{\raggedright\scriptsize\tt#1}}
\def\draft{\oddsidemargin -.2truein
        \def\@oddfoot{\sl preliminary draft \hfil
        \rm\thepage\hfil\sl\today\quad\militarytime}
        \let\@evenfoot\@oddfoot \overfullrule 3pt
        \let\label=\draftlabel
        \let\marginnote=\draftmarginnote
   \def\@eqnnum{(\theequation)\rlap{\kern\marginparsep\tt\@eqnlabel}%
\global\let\@eqnlabel\@vacuum}  }
\def\thebibliography#1{
\vskip 0.5cm \centerline{\bf \Large References}
\list{
[\arabic{enumi}]}{\settowidth\labelwidth{[#1]}
\leftmargin\labelwidth
\advance\leftmargin\labelsep
\usecounter{enumi}}
\def\newblock{\hskip .11em plus .33em minus .07em}
\sloppy\clubpenalty4000\widowpenalty4000
\sfcode`\.=1000\relax}

\renewcommand{\theequation}{\arabic{section}.\arabic{equation}}
\renewcommand{\section}{\setcounter{equation}{0}\@startsection
{section}{1}{0mm}{-\baselineskip}{0.5\baselineskip} {\normalfont\Large\bfseries}}
\renewcommand{\subsection}{\@startsection
{subsection}{2}{0mm}{-\baselineskip}{0.5\baselineskip} {\normalfont\large\bfseries}}
\renewcommand{\subsubsection}{\@startsection
{subsubsection}{3}{0mm}{-\baselineskip}{0.5\baselineskip}
{\normalfont\normalsize\slshape}}

\begin{document}


\begin{titlepage}
\begin{flushright}
CPHT-PC085.092018, September  2018
\vspace{1.5cm}
\end{flushright}
\begin{centering}
{\bm\bf \Large Quantum no-scale regimes and string moduli\footnote{Based on works done in collaboration with T.~Coudarchet and C.~Fleming \cite{CFP,CP}, and presented at the 7th International Conference on New Frontiers in Physics (ICNFP2018).}}

\vspace{5mm}

 {\bf Herv\'e Partouche\footnote{herve.partouche@polytechnique.edu}}

 \vspace{1mm}

{Centre de Physique Th\'eorique, Ecole Polytechnique,  CNRS\footnote{Unit\'e  mixte du CNRS et de l'Ecole Polytechnique, UMR 7644.}, \\ Universit\'e Paris-Saclay, Route de Saclay, 91128 Palaiseau, France}

\end{centering}
\vspace{0.1cm}
$~$\\
\centerline{\bf\Large Abstract}\\

\begin{quote}

\hspace{.6cm} We review that in no-scale models in perturbative string theory, flat, homogeneous and isotropic cosmological evolutions found at the quantum level can enter into ``quantum no-scale regimes'' (QNSRs). When this is the case, the quantum effective potential is dominated by the classical kinetic energies of the no-scale modulus, dilaton and moduli not involved in the supersymmetry breaking. As a result, the evolutions approach the classical ones, where the no-scale structure is exact. When the 1-loop potential is positive, a global attractor mechanism forces the initially expanding solutions to enter the QNSR describing a flat, ever-expanding universe. On the contrary, when the potential can reach negative values, the internal moduli induce in most cases some kind of instability of the growing universe. The latter stops expanding and eventually collapses, unless the initial conditions are tuned in a tiny region of the phase space. Finally, in QNSR, no gauge instability takes place, regardless of the details of the potential.
\end{quote}




\end{titlepage}
\newpage
\setcounter{footnote}{0}
\renewcommand{\thefootnote}{\arabic{footnote}}
 \setlength{\baselineskip}{.7cm} \setlength{\parskip}{.2cm}

\setcounter{section}{0}


\section{Introduction}
\label{intro}

In quantum field theory, what is considered classically is a flat and static universe, with arbitrary particle content. Quantum corrections of the associated fields are then computed, without wondering whether the particle content imposes contraints for the flat universe to be stable.  In fact, for spacetime to remain (nearly) flat and static in presence of gravity, the minima of the quantum potential should (nearly) vanish, a fact that is obviously satisfied but of limited phenomenological interest when supersymmety is exact. On the contrary, when supersymmetry is spontaneously broken, one may look for models at weak string coupling where the 1-loop potential (nearly) vanishes, $\langle \Vone\rangle\simeq 0$ \cite{Itoyama:1986ei, SNSM,L=0,Faraggi}.\footnote{In the framework described in the core of this review, ``nearly vanishes'' means exponentially suppressed in $\Ms/M_{\sigma}$, where $\Ms$ is the string scale and $M_{\sigma}$ the supersymmetry breaking scale measured in $\sigma$-model frame.} Additional conditions should then be imposed for loop corrections up to some higher order to be (nearly) vanishing as well \cite{L2}. If such models with realistic particle content would exist, selecting them may necessitate some version of the anthropic principle. 

In the present paper, we follow a different approach, where the 1-loop effective potential is generic. In this case, the quantum potential energy appears as a source term to be added in the classical Einstein equations. The static universe being no more a solution, we look for the flat, homogeneous and isotropic cosmological evolutions. This is done in the framework of the initially maximally supersymmetric heterotic string in $d$ dimensions, where all supersymmetries are spontaneously broken by a stringy version \cite{SSstring} of the Scherk-Schwarz mechanism \cite{SS}. In this framework, analytic derivations can be done explicitly, allowing us to address two questions \cite{CP,CFP} : 

$(i)$ {\em Under which conditions a flat expanding universe keeps on growing ?} A direct consequence of Einstein equations is that a sufficient condition for this to arise in a perturbative regime is to have $\Vone\ge 0$. It turns out that in the heterotic setup we analyze, the universe can also be ever-expanding when $\Vone <0$. However, the presence of moduli fields often induces some kind of ``instability''~: Most of the initially growing universes stop expanding, and  eventually collapse into Big Crunches. This happens unless the initial conditions are tuned in a tiny region of the phase space. These observations may suggest that spectra leading to positive quantum potentials are way more natural to describe flat expanding spacetimes. 

$(ii)$ {\em What are the properties of these flat, expanding universes ?} In the stringy Scherk-Schwarz mechanism, the supersymmetry breaking scale is a field, $M_{\sigma}\equiv e^{\alpha\Phi}\Ms $ for some constant  $\alpha$,  that parameterizes at tree level a flat direction of a classical potential $\V_{\rm tree}\ge 0$. As a result, $\Phi$  is referred as the ``no-scale modulus'' and the classical model is said to be a ``no-scale model'' \cite{noscale}. In the cosmological evolutions we find at the quantum level, the classical kinetic energies of $M_{\sigma}$, of the dilaton and of moduli fields to be characterized dominate over $|\Vone|$ and the kinetic energies of the remaining moduli. As a result, the expanding solutions approach asymptotically those found classically, when the 1-loop potential is not taken into account, or alternatively when $\Vone$ vanishes (up to exponentially suppressed terms). In other words, the no-scale structure present at the classical level and broken by quantum effects is restored cosmologically. For this reason, we will say that  the universe enters dynamically in ``quantum no-scale regime'' (QNSR).  Moreover, we will see that no Higgs-like instability occurs in such a regime, whether moduli sit at minima, maxima, saddle points or actually anywhere in moduli space. 

In Sect.~\ref{toy}, we describe a toy model involving a minimal set of degrees of freedom, in order to highlight the existence of QNSRs \cite{CFP}. The effects of marginal deformations in true heterotic models are described in Sect.~\ref{full} \cite{CP}. Our concluding remarks can be found in the last section. 


\section{Toy model}
\label{toy}

The positive thing in considering a reduced set of degrees of freedom is that solving exactly the equations of motions at 1-loop is possible. To be specific, we consider the heterotic string with classical background
\be
\label{fac}
\R^{1,d-1}\times \prod_{i=d}^{d+n-1}S^1(R_i)\times T^{10-d-n}\, ,
\ee
where the size of $T^{10-d-n}$ is of the order of the string scale. A coordinate-dependent compactification~\cite{SSstring} along the $n$  circles of radii $R_i$ is implemented, which is the counterpart in string theory of the Scherk-Schwarz mechanism in supergravity \cite{SS}. The radii $R_i$ are chosen large (compared to 1) for the supersymmetry breaking scale to be lower than the string scale,
\be
M_{\sigma}\equiv{\Ms \over \big( R_d\cdots R_{d+n-1}\big)^{1\over n}}\ll \Ms\, .
\ee 
This restriction ensures that no Hagedorn instability can occur, and allows a great simplification of the expression of the 1-loop effective potential. The latter, defined in terms of the partition function $Z$, takes the following form \cite{Itoyama:1986ei,SNSM}
\be
\label{vs}
\Vone^{\sigma}\equiv -{\Ms^d\over (2\pi)^d}\int_\F {d^2\tau\over 2\tau_2^2}\, Z= (\nF-\nB)\, v_{d} \, M_{\sigma}^d +\O\left((\Ms M_{\sigma})^{d\over 2}\, e^{-\Ms/M_{\sigma}}\right),
\ee
where $\nB,\nF$ are the numbers of massless bosons and fermions.\footnote{As usual, $\tau=\tau_1+i\tau_2$ is the genus-1 Teichm\"uller parameter and $\F$ the fundamental domain of $SL(2,\Z)$.} The dominant contribution of this expression arises from the towers of Kaluza-Klein (KK) modes associated to the large supersymmetry breaking circles. In fact, $v_d$ is a function of the complex structures $R_i/R_d$, we will treat in this section as constants. The contributions of all other string states are exponentially suppressed and will be neglected from now on. At 1-loop order, the effective action involves the classical kinetic terms of $M_\sigma$ and of the dilaton $\phi_{\rm dil}=\langle \phi_{\rm dil}\rangle+\phi$, as well as the potential.\footnote{When $\Vone$ does not vanish identically, it is a source for the kinetic energies. Adding 1-loop corrections to the kinetic terms would effectively modify the solutions at second order in string coupling.} In Einstein frame, we obtain 
\be
S={1\over \kappa^2}\int d^dx  \sqrt{-g}\, \bigg[{{\cal R}\over 2}-{1\over2}(\partial\Phi)^2-{1\over 2}(\partial\phi_\bot)^2-\kappa^2\Vone\bigg] ,
\ee
where we have introduced canonical scalars $\Phi$ and $\phi_\bot$. The former is the no-scale modulus, which is related to the supersymmetry breaking scale measured in this frame, while $\phi_\bot$ is an ``orthonormal'' combination of fields, 
\begin{align}
M&=e^{{2\over d-2}\phi} M_\sigma=e^{\alpha \Phi}\Ms\, , \;\; \where\;\; \alpha=\sqrt{{1\over d-2}+{1\over n}} \, ,\espD\nonumber \\
\phi_\bot&={1\over  \sqrt{d-2+n}} \Big[2 \phi+\ln \Big(R_d\cdots R_{d+n-1}\Big)\Big].
\end{align}
In our notations, ${\cal R}$ is the Ricci curvature, $\kappa^2=e^{2\langle \phi_{\rm dil}\rangle}/\Ms^{d-2}$ is the Einstein constant and the potential becomes
\be
\label{Va}
\Vone=  (\nF-\nB)\, v_{d}\,e^{d\alpha\Phi} \Ms^d \, .
\ee
Ignoring higher order corrections in string coupling, the field configurations which make extremal this action are relevant when compatible with weak coupling. 

\begin{figure}[h]
\centering
\includegraphics[width=7.0cm]{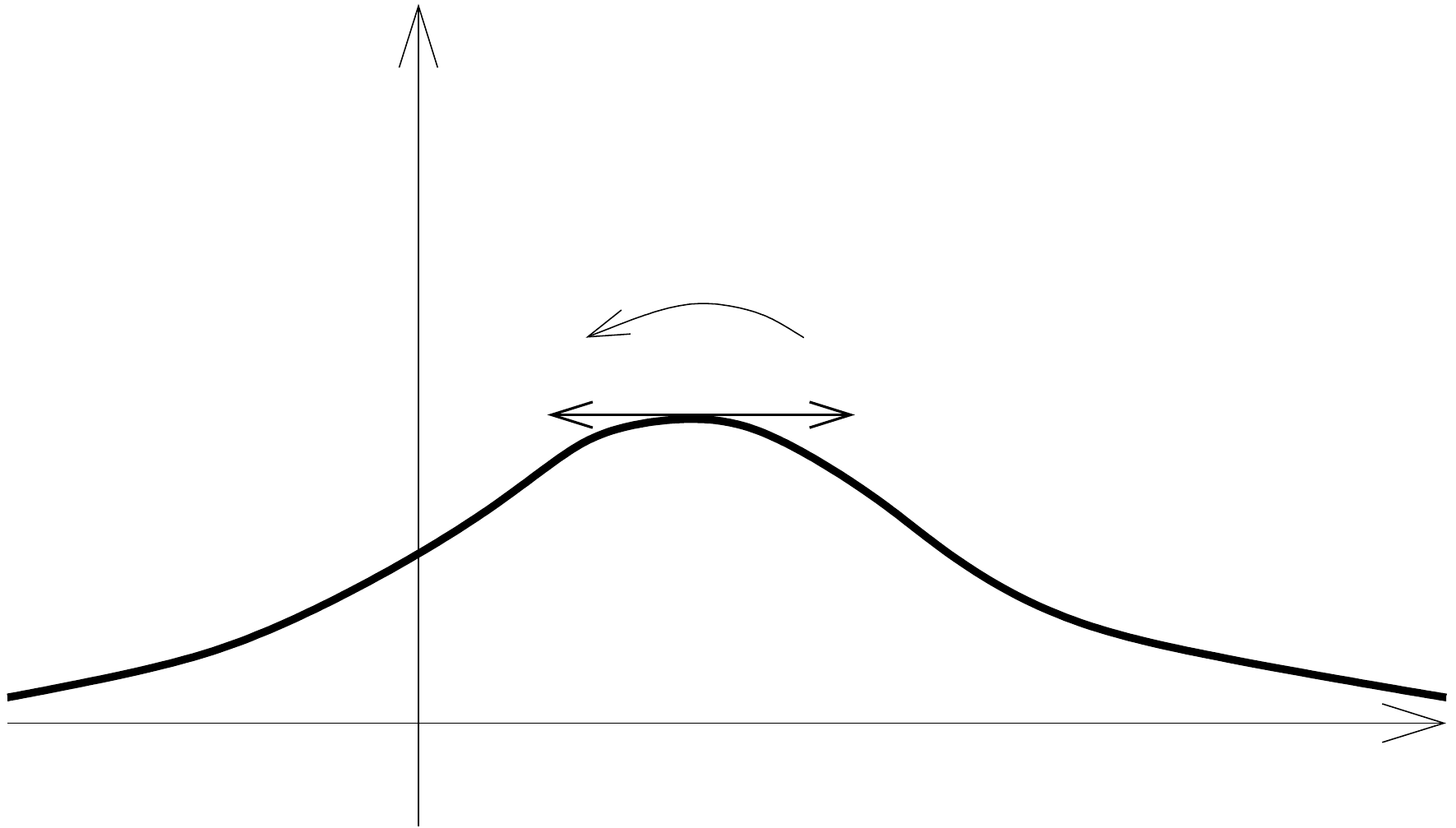}\qquad \;
\includegraphics[width=7.0cm]{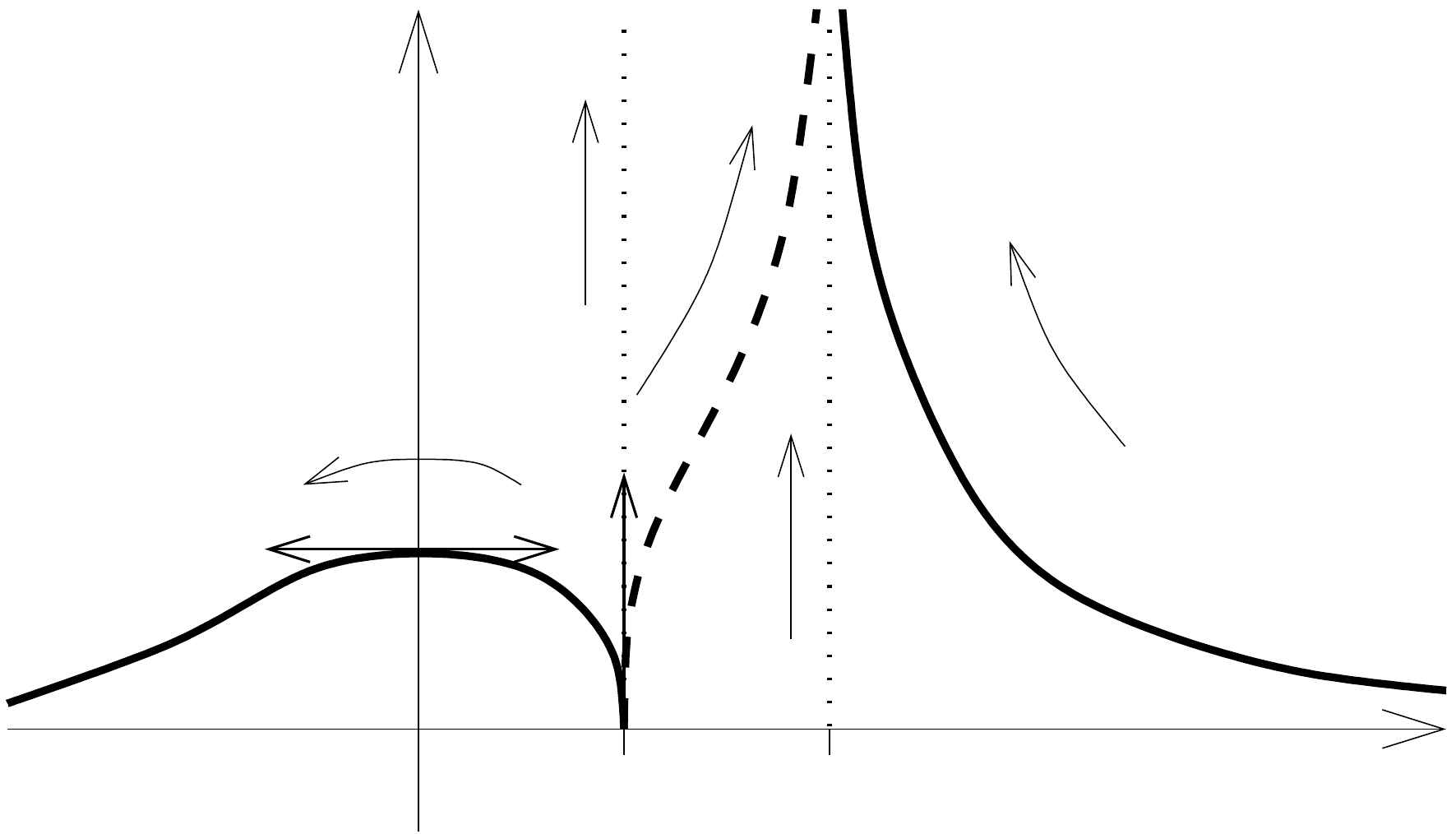}
\begin{picture}(0,0)
\put(-241,1){$\tau$}\put(-11,1){$\tau$}
\put(-388,107){$a$}\put(-158.5,107){$a$}
\put(-433,99){\mbox{\Large$(i)$}}\put(-203.5,99){$\mbox{\Large$(ii)$}$}
\put(-121,1){$\tau_-$}\put(-93,1){$\tau_+$}
\end{picture}
\caption{\footnotesize Qualitative behaviors of $a(\tau)$, $(i)$~in the supercritical and $(ii)$~subcritical cases. The arrows indicate the directions of the evolutions for increasing cosmic time $t$, when $c_\Phi>0$. Solid and dashed curves correspond to no-scale models with $\nF-\nB<0$ and $\nF-\nB>0$, respectively. Dotted lines describe the classical trajectories.}
\label{fig_a}
\end{figure}

Being interested in homogeneous and isotropic cosmological evolutions in flat space, we consider the scalar fields $\Phi(t)$, $\phi_\bot(t)$ and metric $g_{\mu\nu}={\rm diag}(-1,a(t)^2,\dots,a(t)^2)$ ansatz, where $t$ is cosmic time. If $\phi_\bot$ is a free field, it turns out that the proportionality of $\partial \Vone/\partial \Phi$ and $\Vone$ leads to the existence of a second free field, which involves the no-scale modulus. Integrating once their  equations of motion, we obtain
\be
\dot \phi_\bot=\sqrt{2}\, {c_\bot\over a^{d-1}}\, , \qquad \alpha\dot\Phi+ {\alpha^2\over 2} d(d-2)  H= {c_\Phi\over a^{d-1}}\, , 
\ee
where $c_\bot$ and $c_\Phi$ are constants. These results can be used in the equation obtained by varying $S$ with respect to $a$,
\be
{1\over 2}\, (d-2)\, \dot H=-{1\over2}\, \dot\Phi^2-{1\over 2}\, \dot\phi_\bot^2\, , 
\label{eqq}
\ee
where $H=\dot a/a$, to obtain a second order differential equation in scale factor  only. The latter can be  solved when $c_\Phi\neq 0$  by defining a new (dimensionless) time variable $\tau$, in terms of which Eq.~(\ref{eqq}) becomes
\begin{align}
&A\, {da\over a}=-{\tau d\tau\over \P(\tau)}\, ,  \qquad \where \qquad \tau= {2A\over d(d-1)c_\Phi}\, (a^{d-1})^{\displaystyle\cdot}\, , \nonumber \\
&\P(\tau)=\tau^2-2\tau+\omega\, \Big[1+2\alpha^2\Big({c_\bot\over c_\Phi}\Big)^2\Big]\, ,   \quad A={\omega\over 4}\, d^2(d-2)\alpha^2 \, , \quad  \omega=1-{4(d-1)\over d^2(d-2)\alpha^2}\, .
\end{align}
Finally, it is convenient to determine the second integration constant of the no-scale modulus by using Friedmann equation, which takes an algebraic form in terms of $\tau$,
\be
(\nF-\nB)\, v_d\,\kappa^2M^d=-{c_\Phi^2\over 2\alpha^2\omega}\, {\P(\tau)\over a^{2(d-1)}}\, .
\label{fri}
\ee
In the following, we present the cosmological solution found for arbitrary $c_\bot/c_\Phi$, for which a critical value turns out to be \cite{CFP}
\be
\gamma_c=\sqrt{1-\omega\over 2\alpha^2\omega}\, .
\ee
\paragraph{\large Spercritical case, \bm  $\displaystyle {c_\bot\over \gamma_c c_\Phi}>1$ : } The degree 2 polynomial $\P(\tau)$ has no  real root. Due to Friedmann equation~(\ref{fri}), this case arises only in models where $\nF-\nB<0$. Moreover, no classical limit $\kappa^2\to 0$ exists, and the evolutions are thus  intrinsically of quantum nature.  The scale factor, which involves an integration constant $a_0$,  is found to be
\be
\label{sol1}
a=a_0\,  {e^{-{1\over As }\arctan({\tau-1\over s  })}\over \P(\tau)^{1\over 2A}}\, 
\, , \quad \where \quad  s=\sqrt{1-\omega}\; \sqrt{\Big({c_\bot\over \gamma_cc_\Phi}\Big)^2-1}\, ,
\ee 
and is shown in Fig.~\ref{fig_a}($i$). The arrow indicates the direction of the evolution when $t$ increases and $c_\Phi >0$. 
Translating the $\tau\to \pm \infty$ limits in cosmic time, the scale factor behaves as $a(t)\sim \#\, |t-t_\pm|^{1\over A+d-1}$, for some constants $t_\pm$. In these limits, the evolution describes a Big Bang and a Big Crunch, where the universe is dominated by the total energy of $M$, \ie kinetic plus quantum potential, 
\be
H^2\sim\#\,  \dot\Phi^2\sim\#\,  \kappa^2\Vone\sim  {\#\over a^{2(A+d-1)}}\gg \dot\phi_\bot^2={\#\over a^{2(d-1)}}\, .
\ee
The 1-loop potential is also responsible for stopping the expansion of the universe, since we have $\ddot a\propto \Vone<0$ at the maximum of $a$. In fact, the evolution can be trusted far enough from the Big Bang and the Big Crunch, for several reasons: First, all kinetic energies must be bounded by the string scale, a fact that is guaranteed as far as $a(t)$ is not too low. Second, the string coupling must remain weak, which imposes a limitation on the range of $\tau$, since we have  $e^{2d\alpha^2\phi}\sim \# |\tau|^{2\over \omega}$, as $\tau\to \pm \infty$.  To conclude on the supercritical case, the  generation of negative quantum potentials allows new universes to exist (they have no classical counterparts), however ``unstable'' in the sense that the potentials are also responsible for their collapses.  

For completeness, we signal that the case $c_\Phi=0$ is somehow ``infinitely supercritical'' when $c_\bot\neq 0$. It requires $\nF-\nB<0$ and yields an evolution qualitatively similar to the above  one. When $c_\Phi=c_\bot=0$, the maximum of the scale factor is formally sent to infinity, and the trajectories become monotonic. 
   
\paragraph{\large \bm Subcritical case, $\displaystyle{c_\bot\over \gamma_c c_\Phi}<1$ :} The polynomial $\P(\tau)$ has 2 distinct real roots,
\be
\tau_\pm=1\pm r\, , \quad \where\quad r=\sqrt{1-\omega}\; \sqrt{1-\Big({c_\bot\over \gamma_cc_\Phi}\Big)^2}\, .
\label{re}
\ee
From Eq.~(\ref{fri}), we see that there is no restriction on the massless spectrum of the models  : 

$\bullet$ When  $\nF-\nB>0$, the range of allowed time is $\tau_-<\tau<\tau_+$.   

$\bullet$ When $\nF-\nB=0$, $\tau(t)$ is constrained to be identically equal to $\tau_+$ or $\tau_-$. 

$\bullet$ When $\nF-\nB<0$, two branches of solutions are allowed, namely $\tau<\tau_-$ and   $\tau_+<\tau$.

Let us first describe the case $\nF-\nB=0$. In the literature, the models with massless spectra satisfying this condition are sometimes referred as  ``super no-scale models''. This follows from the fact that at the points in moduli space associated to the classical backgrounds, the 1-loop effective potential is extremal (with respect to all internal space moduli), vanishes and is degenerate for arbitrary value of the no-scale modulus.\footnote{Up to exponentially suppressed terms.\label{up}} Thus, when the extremum is a minimum, the exact classical no-scale structure is promoted to the 1-loop level.${}^{\scriptsize\mbox{\ref{up}}}$ When the 1-loop effective potential does not contribute at all in the action $S$, one should in principle add the 1-loop corrections to the kinetic terms of $\Phi$ and $\phi_\bot$. However, once marginal deformations associated to  the internal space are taken into account, as is done in Sect.~\ref{full}, $\Vone$ is not identically vanishing even when $\nF-\nB=0$. Thus, the present case where the potential is formally absent does not need to be considered in very details and we  only present the limit behaviors of the 1-loop evolutions. In perturbative regime, they can be  found easily  by extremizing the  classical action. For $\tau\equiv \tau_\pm$, we find the expanding (or contracting, by time reversal) behaviors
\be
H^2\sim \# \,\dot\phi_\bot^2  \sim \#\,  \dot\Phi^2\sim  {\#\over a^{2(d-1)}}\, ,\quad  \where \quad a\sim \# |t-t_\pm|^{1\over d-1}\, , 
\ee
which imply in particular 
\be
M^d\sim {\#\over a^{2(d-1)+K_\pm}}\, , \quad \where \quad K_\pm=\pm {2Ar\over 1\pm r}\, . 
\label{kpm}
\ee
It can be checked that for the evolution $\tau\equiv \tau_+$, an interval of $c_\bot/c_\Phi$ exists such that $\phi(t)\to -\infty$ as $c_\Phi(t-t_+)\to +\infty$, which  yields  a consistent perturbative expanding universe. Similarly, for    $\tau\equiv \tau_-$, the Big Crunch behavior $c_\Phi(t-t_-)\to 0_+$ is also perturbative in some finite range of $c_\bot/c_\Phi$ \cite{CFP}.

In generic models, where $\nF-\nB\neq 0$, the scale factor and supersymmetry breaking scale evolutions are found to be 
\be
a={a_0\over | \tau-\tau_-|^{1\over K_-}\, |\tau-\tau_+|^{1\over K_+}}\, , \quad 
M^d={M^d_+\over a^{2(d-1)+K_+}} \left|{\tau-\tau_-\over \tau_+-\tau_-}\right|^{2\over \tau_+}={M^d_-\over a^{2(d-1)+K_-}} \left|{\tau_+-\tau\over \tau_+-\tau_-}\right|^{2\over \tau_-},
\label{aM}
\ee
where $M_\pm$ can be expressed in terms of $c_\Phi$, $\tau_+$ and $\tau_-$. As announced before and shown in Fig.~\ref{fig_a}($ii$), the expression of $a(\tau)$ admits two branches when $\nF-\nB<0$ (in solid lines), while for $\nF-\nB>0$ a unique branch exists (in dashed line). The classical trajectories $\tau\equiv \tau_\pm$ are also plotted (in dotted lines). Some remarks are in order : 

 $\bullet$ All trajectories start and/or end with a Big Bang or Big Crunch.   

 $\bullet$ The behavior of the solutions as $\tau\to\pm \infty$, which exists when $\nF-\nB<0$,  is identical to that found in the supercritical case.  

 $\bullet$ More interesting is the fact that all solutions approach $\tau_+$ and/or $\tau_-$ \ie the classical evolutions, which are also the limit behaviors found in the super no-scale case $\nF-\nB=0$. To be specific, when $\tau\to \tau_\pm$ we have
\be
H^2\sim \# \, \dot \phi_\bot^2\sim \#\,  \dot\Phi^2\sim \, {\#\over a^{2(d-1)}} \gg |\Vone| \sim {\#\over a^{2(d-1)+K_\pm}}\, , 
\ee
which shows that the no-scale structure is restored during the cosmological evolution. By definition, we will say that the universe enters ``quantum no-scale regimes'', which are characterized by the no-scale modulus $\Phi$ becoming free. 

 $\bullet$ The conditions for the QNSRs to be compatible with string weak coupling are identical to those found in the classical case.  

 $\bullet$ When $\Vone>0$, the initially expanding solutions are always attracted to the ever-expanding QNSR $\tau\to+\infty$. However, in the models where $\Vone <0$, only one out of the two branches describes this behavior, while the second branch yields initially expanding universes sentenced to collapse. These two drastically different histories are somehow on equal footing, depending on the choice of initial conditions for $\tau$, greater than $\tau_+$ or lower than $\tau_-$. This very fact turns out to be amended when the dynamics of other moduli fields is included, as will be seen in the next section. 

 $\bullet$ The supersymmetry breaking scale admits a non-trivial dynamic. In the branch $\tau>\tau_+$, where the potential is negative, Eq.~(\ref{aM}) shows that when the universe expands ($\tau\to \tau_+$),  $M$ decreases. Thus, the no-scale modulus climbs its exponential potential.  In addition, when $\tau$ varies between $\tau_-$ and $\tau_+$, $M$ climbs and then descends its positive potential, when the integration constant satisfy $|c_\bot/(\gamma_c c_\Phi)|<\sqrt{\omega}$ (see also \cite{DKS}). Around the maximum of $M$, the cosmological evolution may accelerate but with at most an $e$-fold number of order 1, thus not describing a substantial inflation. 

For completeness, we mention that the critical case, namely $|c_\bot/(\gamma_c c_\Phi)|=1$, for which the polynomial $\P(\tau)$ admits a double root $\tau_+=\tau_-=1$, is qualitatively similar to the subcritical one, up to the fact that it applies only to models where $\nF-\nB\le  0$. 


\section{Including moduli fields}
\label{full}

Having defined the QNSRs in a toy model, our aim is to show how they are affected by the dynamics of marginal deformations \cite{CP}. Compactifying the heterotic string on $T^{10-d}$, the Narain moduli space 
\be
{SO(10-d,26-d)\over SO(10-d)\times SO(26-d)}
\ee
can be parameterized by the internal metric and antisymmetric tensor $G$, $B$, as well as $Y$-fields. They can be split into constant backgrounds and $y$-deformations,
\be
(G+B)_{Ij}={(G^{(0)}+B^{(0)})_{Ij}}+\sqrt{2}\, y_{Ij}\, , \;\;I,j\in\{d,\dots,9\}\, , \quad Y_{Ij}=Y_{Ij}^{(0)}+y_{Ij}\, ,  \;\;j\in\{10,\dots,25\}\, ,
\ee
where $y_{Ij}$ is the Wilson line of the $j$-th Cartan $U(1)$ along $X^I$. For simplicity, we implement the coordinate-dependent compactification along $n=1$ direction, $X^d$. As a result, the supersymmetry breaking scale in $\sigma$-model frame is $M_\sigma=\sqrt{G^{dd}} \Ms$. As before, we take the direction $X^d$ to be large, while the size of the remaining compact coordinates is of the order of the string scale. 

Once we take into account marginal deformations, $\nF$ and $\nB$ are not strictly speaking integer constant. This is due to the fact that they  interpolate between distinct integer values valid in different regions of moduli space. Choosing an initial background where no non-zero mass scale exists below $\Ms$, the 1-loop effective potential~(\ref{Va}) now involves $\nF$ and $\nB$ that are functions of the $y$-deformations. Taylor expanding, one obtains \cite{SNSM, CP}
\be
\Vone=(\nF^{(0)}-\nB^{(0)})\,  v_{d}\,  M^d + M^d \, {v_{d-2}\over 2\pi}\! \sum_{j=d+1}^{25}\!\big(C^{j}_{\rm B}-C^{j}_{\rm F}\big)\Big[(d-1)y^2_{dj}+{1\over G^{dd}}\big(y^2_{d+1,j}+\cdots+y^2_{9j}\big)\Big]+\cdots,
\ee
where $C_{\rm B}^j$ ($C_{\rm F}^j$) is ${1\over 2}$ times the sum of the (charges)$^2$ of all massless bosons (fermions) of the undeformed background, with respect to the $j$-th Cartan $U(1)$. As a result, Wilson lines may be massive, massless or tachyonic. 
To analyze all cases, we need to consider backgrounds with factors $C^{j}_{\rm B}-C^{j}_{\rm F}$ of arbitrary signs. Let us first remind that in the most standard implementation of the Scherk-Schwarz mechanism, the massless bosonic (fermionic) fields in $d+1$ dimension are periodic (antiperiodic)  along $X^d$. As a result, denoting $m_d\in\Z$ and $F\in\Z_2$ the momentum and  fermionic number of the descendent modes in $d$ dimensions, the generalized momentum along $X^d$ is 
\be
P_d\equiv m_d+{1\over 2}\, F+\cdots+(G+B)_{dj}n_j= m_d+{1\over 2}\Big[ F+2(G^{(0)}+B^{(0)})_{dj}n_j \Big]+\cdots+\sqrt{2}\, y_{dj}n_j \, , 
\ee
where $n_j\in\Z$ is the winding number along $X^j$, $j\in\{d+1,\dots,9\}$. In this formula, the ellipses stand for irrelevant contributions involving the $Y$-fields. When the  background is chosen so that $2(G^{(0)}+B^{(0)})_{dj}n_j$ is an even integer, the lightest bosons are massless at $y_{dj}=0$, while the lightest fermions have a KK mass. However,  when $2(G^{(0)}+B^{(0)})_{dj}n_j$ is an odd integer, the role of bosons and fermions is reversed : The lightest bosons have a KK mass at $y_{dj}=0$, while the lightest fermions are massless.

As an example to be detailed in great details until the end of this paper, consider the matrix $(G+B)_{Ij}$ for $I,j\in\{d,d+1\}$, with $b\in\Z$,
\be
(G+B)_{Ij}=\begin{pmatrix}R_d^2 & \displaystyle{b\over 2}+\sqrt{2}\, y_{d,d+1}\\  -\displaystyle{b\over 2} +\sqrt{2}\, y_{d+1,d}& 1+\sqrt{2}\, y_{d+1,d+1}\end{pmatrix}  .
\ee
When $R_d\to +\infty$ and supersymmetry is recovered in $d+1$ dimensions, 2 vector multiplets with $m_{d+1}=-n_{d+1}=\pm 1$ are massless when $b$ is even and the $y$-deformations vanish. As a result, the Cartan $U(1)$ gauge symmetry arising from the direction $X^{d+1}$ is enhanced to $SU(2)$. When $R_d$ is finite, the fermionic parts of these vector multiplets acquire a mass $M_\sigma$, the $SU(2)$ gauge symmetry remains valid and we end up with a mass coefficient $C_B^{d+1}-0=8\times 2$. However, $R_d$ finite allows the other interesting case where  $b$ is odd. In this case, the fermionic parts of the vector multiplets are massless, the gauge symmetry becomes $U(1)$ and the mass coefficient is   $0-C_F^{d+1}=-8\times 2$. In fact, freezing for simplicity the degrees of freedom
\be
y_{I,d+2}\equiv \dots\equiv y_{I,25}\equiv 0, \; I\in\{d,\dots,9\}\, , \qquad y_{id}\equiv y_{i,d+1}\equiv 0, \; i\in\{d+2,\dots,9\}\, , 
\label{rest}
\ee
the 1-loop effective potential for arbitrary $y_{d,d+1}$, $y_{d+1,d}$ and $y_{d+1,d+1}$ turns out to be, in $\sigma$-model frame,
\begin{align}
&\Vone^{\sigma}= \big(\nF^{(0)}-\nB^{(0)}+(-1)^b\, 8\times 2\big)\,  v_{d}\,  M_{\sigma}^d\nonumber \espD\\
&\;\; \;- (-1)^b\, 16\, {2M_{\sigma}^d\over (2\pi)^{3d+1\over 2}}   \sum_{k} {\cos \big(2\pi( 2k+1) z\big)\over |2k+1|^{d+1}} \, {\cal H}\bigg(2\pi |2k+1| {{\cal M}\over M_{\sigma}}\bigg)+O\!\left((\Ms M_{\sigma})^{d\over 2}\, e^{-\Ms/M_{\sigma}}\right),\nonumber\espDD \\
&\where \;\;\;  \M={\sqrt{2}\, |y_{d+1,d+1}|\Ms\over \sqrt{1+\sqrt{2}\,y_{d+1,d+1}}}\, , \;\; \;z=\sqrt{2}\, \bigg(y_{d,d+1}-{y_{d,d+1}+y_{d+1,d}\over \sqrt{2}\left(1+\sqrt{2}\, y_{d+1,d+1}\right)}\, y_{d+1,d+1}\bigg) .
\label{po}
\end{align}
In this result, $\M$ characterises how the $SU(2)$ non Cartan fields are massive. When they are heavy, $\M>M_\sigma$, all of the second line of $\Vone^\sigma$ is exponentially suppressed, as follows from the function ${\cal H}(x)\equiv x^{d+1\over 2}K_{d+1\over 2}(x)$. On the contrary, when $\M<M_\sigma$, the potential has a simple $U$-shape, with a minimum at $\M=0$, when $b$ is even and $z=0$. Moreover, the potential is 1-periodic in $z=\sqrt{2}\, y_{d,d+1}+\cdots$ and the second line evolves essentially as $\cos(2\pi z)$. Consistently, changing $b$ even to $b$ odd amounts to shifting $z\to z+{1\over 2}$. Note that since there are only 2 combinations $\M$ and $z$ of Wilson lines on which the potential depends, the latter admits a flat direction $\sqrt{2}\, y_{d+1,d}+\cdots$ \cite{CP}. 

\paragraph{\large Small Wilson line deformations :}

In order to show the existence of QNSRs in presence of dynamical Wilson lines, it is enough to consider the case of small deformations, $|y_{d,d+1}|\ll 1$, $|y_{d+1,d}|\ll 1$, $|y_{d+1,d+1}|\ll \sqrt{G^{dd}}$. At quadratic order in $y$'s, the 1-loop effective action to be considered is 
\begin{align}
S={1\over \kappa^2}\int d^dx \sqrt{-g}\, \bigg[&{{\cal R}\over 2}- {1\over 2}(\partial\Phi)^2-{1\over 2} (\partial\phi_\bot)^2  -{1\over 4} (\partial y_{d+1,d+1})^2\nonumber \\
&-{G^{dd}\over 4}(\partial y_{d,d+1})^2-{G^{dd}\over 4}(\partial y_{d+1,d})^2+\cdots-\kappa^2\Vone\bigg] .
\end{align}
The ellipses stand for higher order corrections in Wilson lines we neglect for the time being, $G^{dd}=e^{{2\over\alpha} \Phi}\, e^{-{2\over \sqrt{d-1}}\phi_\bot}$, and the potential is 
\be
\Vone =   e^{d\alpha\Phi} \Ms^d \bigg[\big(\nF^{(0)}-\nB^{(0)}\big)\,  v_{d}+ (-1)^b\, {8\over \pi}\, v_{d-2} \, \Big(\!(d-1)y^2_{d,d+1}+{y^2_{d+1,d+1}\over G^{dd}}\Big)\bigg]\!+\cdots.
\ee
After deriving the equations of motion, evolutions describing flat, homogeneous, and isotropic ever-expanding universes or Big Bangs are sought, with 1-loop potential dominated in Friedmann equation~: 
\be
a(t)\underset{t\to +\infty}{\longrightarrow} +\infty \; \;\mbox{ or } \; \; a(t)\underset{t\to t_-}{\longrightarrow} 0\, , \qquad \kappa^2 \Ms^d\, e^{d\alpha \Phi} =\O\left({H^2\over a^{K_\pm}}\right) .
\label{li}
\ee
In the above ansatz, $\pm K_\pm>0$ is a constant to be determined by consistency. We find the following : 

 $\bullet$
The simplest behavior to derive is that of the scale factor, $a\sim \#|t-t_\pm|^{1\over d-1}$, which is a consequence of the negligible potential. 

 $\bullet$  As can be seen from the action $S$, the Wilson lines $y_{d,d+1}$ and $y_{d+1,d}$ have non-trivial friction terms. Moreover, the positive or negative mass term of $y_{d,d+1}$ turns out to be irrelevant in the limits~(\ref{li}), so that 
\be
\dot y_{d+1,d}= {\#\over a^{d-1}G^{dd}}\; , \qquad \dot y_{d,d+1}\sim {\#\over a^{d-1}G^{dd}}\, .
\ee
An important remark then follows. In Friedmann equation, 
\be
{1\over 2}\, (d-1)(d-2) H^2={1\over2}\, \dot\Phi^2+{1\over 2}\, \dot\phi_\bot^2+{G^{dd}\over 4}\, \dot y_{d,d+1}^2+{G^{dd}\over 4}\, \dot y_{d+1,d}^2+{1\over 4}\, \dot y_{d+1,d+1}^2 +\kappa^2 \Vone\, , 
\ee
the l.h.s. behaves as $H^2\sim \#/(t-t_\pm)^2$, while in the r.h.s. the kinetic energies of $y_{d,d+1}$ and $y_{d+1,d}$ yield a contribution $\sim{\#/[(t-t_\pm)^2G^{dd}}]$, and the potential is dominated. By consistency, we  assume that $G^{dd}\sim \#(t-t_\pm)^{J_\pm}$, where $\pm J_\pm>0$ is a constant to be determined. Under this assumption, one obtains that $y_{d,d+1}$ and $y_{d+1,d}$ converge to arbitrary constants, which are small in the present paragraph. Note that even when its mass term is negative, $y_{d,d+1}$ does not automatically develop a large expectation value, which would destabilize drastically the initial background. 

 $\bullet$ $y_{d+1,d+1}$ has a positive or negative mass term, irrelevant in the limits we consider. As a result, one obtains
\be
\dot y_{d+1,d+1}\sim {2c_y\over a^{d-1}}\quad \Longrightarrow\quad |y_{d+1,d+1}|\sim \#|\ln(t-t_\pm)|\ll \sqrt{G^{dd}}\, , 
\ee
where $c_y$ is an integration constant. Due to the fact that $G^{dd}\sim \#(t-t_\pm)^{J_\pm}$, with $\pm J_\pm>0$, we have $\M/M_\sigma \to 0$. This shows that even if $y_{d+1,d+1}$ has a logarithmic behavior, it is effectively attracted to the extremum of the potential at $y_{d+1,d+1}=0$. In other words, whether the potential has a minimum at $\M=0$ (when $b$ is even) or a maximum  (when $b$ is  odd), the theory is always approaching the configuration where $\M$ effectively vanishes. 

 $\bullet$  $\phi_\bot$ and $\Phi$ couple to the kinetic and mass terms of Wilson lines. However,  in the regimes we are interested in, they are becoming free,
\be
\dot\phi_\bot\sim \sqrt{2}\, {c_\bot\over a^{d-1}}\, , \quad \; \alpha\dot\Phi+  {\alpha^2\over 2}\, d(d-2)\,  H\sim {c_\Phi\over a^{d-1}}\, ,
\ee
where $c_\bot$, $c_\Phi$ are constants. 

\begin{figure}[h]
\centering
\includegraphics[width=6.0cm]{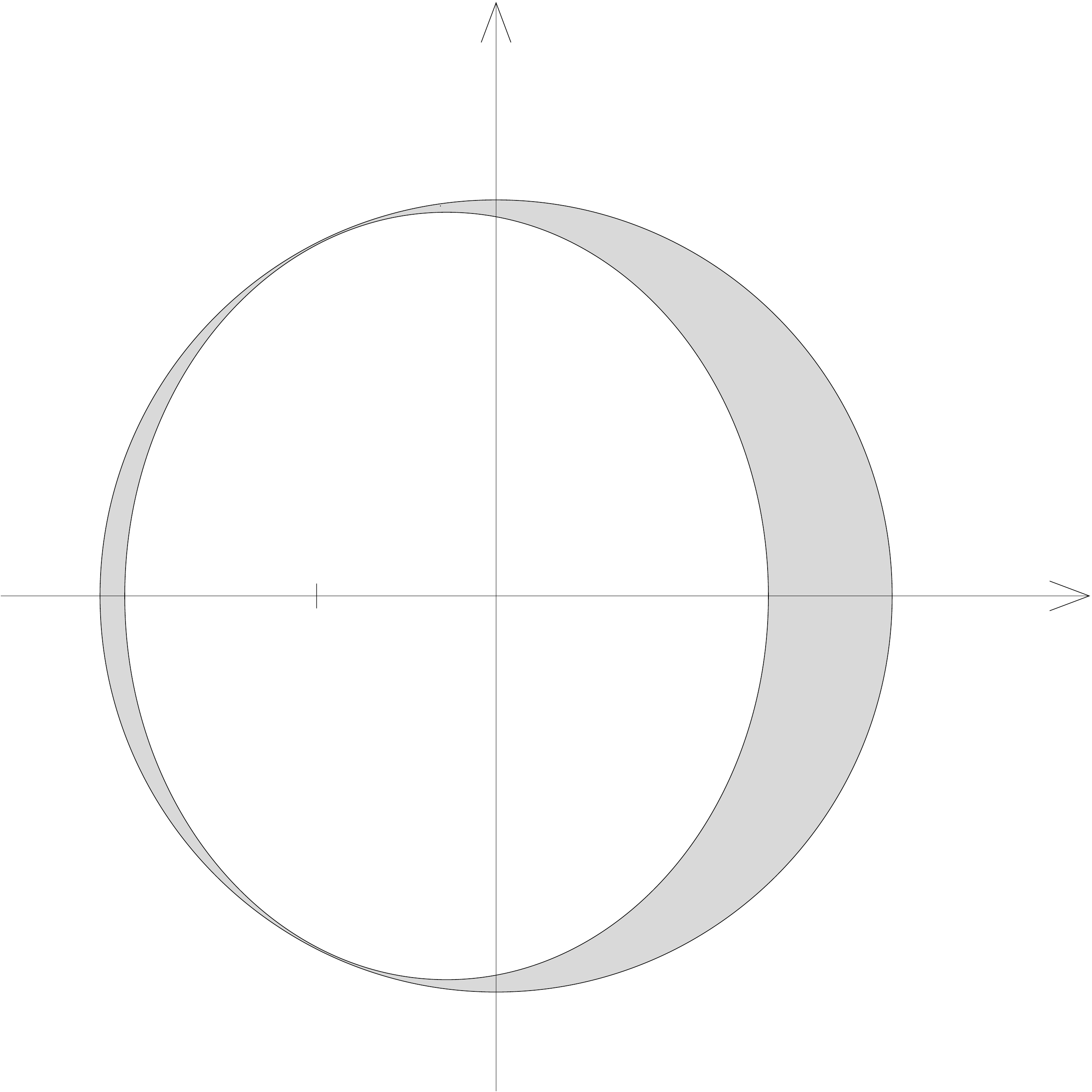}
\begin{picture}(0,0)
\put(-27.5,59){$\displaystyle {c_\bot\over \gamma_{\rm c} c_\Phi}$}
\put(-126,157){$\displaystyle  {c_y\over \gamma_{\rm c} c_\Phi}$}
\put(-37.8,65){1}
\end{picture}
\caption{\footnotesize The points $\big({c_\bot\over \gamma_{\rm c}c_\Phi},{c_{y}\over \gamma_{\rm c}c_\Phi}\big)$ of the disk of  radius 1 that yield QNSRs $a\to\infty$ and $a\to 0$ sit in the left and right shaded crescents, respectively.}
\label{di}
\end{figure}

 $\bullet$  From the above relations, the coefficients $K_\pm$ and $J_\pm$ can be determined in terms of $c_\bot/c_\Phi$ and  $c_y/c_\Phi$. The result for $K_\pm$ is that found in the toy model, Eq.~(\ref{kpm}), up to the change $c_\bot^2\to c_\bot^2+c_y^2$ in the expression of $r$ given in Eq.~(\ref{re}). As a result, the subcritical region for which QNSRs exist, which is a segment in the previous section, becomes a disk of radius 1, as shown in Fig.~\ref{di}.  
However, the additional condition $\pm J_\pm>0$ arising from the dynamics of $y_{d,d+1}$ and $y_{d+1,d}$ amounts to restricting $\big({c_\bot\over \gamma_cc_\Phi}, {c_y\over \gamma_cc_\Phi}\big)$ to the shaded regions~: The left and right crescents yield QNSRs $a\to +\infty$ and $a\to 0$, respectively. 

 $\bullet$  Due to this restriction to the shaded domains, the QNSR $a\to +\infty$ turns out to  always be perturbative. The $a\to 0$ one is also perturbative, except in the neighborhood of the tips of the right crescent \cite{CFP}. 

\noindent This concludes the proof that QNSRs  exist even when the  dynamics of internal moduli fields is taken into account. However, it is important to stress that in Fig.~\ref{di}, the left crescent is actually much smaller than that we have drown. Its width along the horizontal axis is between $10^{-3}$ and $10^{-2}$ , for $3\le d\le 9$. However, one should not conclude that be reached, QNSRs always imply some kind of fine tuning, as explained in the following.

\paragraph{\large Numerical simulations}

To analyze whether global attractor mechanisms avoid the need to tune integration constants or initial conditions for  the QNSR $a\to \infty$ to be reached, we consider computer simulations. For this purpose, restricting to the spacetime dimension $d=4$, we simulate numerically the quantities
\be
 c^{\rm sim}_{\bot}=\frac{a^{d-1}}{\sqrt{2}}\, \dot \phi_\bot\, , \quad\;  \;c^{\rm sim}_{\Phi} =a^{d-1}\left(\alpha\dot{\Phi}+{\alpha^2\over 2} d(d-2)H\right), \quad \;\; c^{\rm sim}_y =\frac{ a^{d-1}}{2}\, \dot y_{55}\, , 
\ee
which are expected to converge to constants $c_\bot$, $c_\Phi$, $c_y$. In fact, we have checked that if $\Big({c^{\rm sim}_\bot\over \gamma_cc^{\rm sim}_\Phi}, {c_y^{\rm sim}\over \gamma_cc_\Phi}\Big)$ starts in the left thin shell of Fig.~\ref{di} and that the initial velocities are low, the above quantities do freeze. Moreover, the climbing of $M$ when its potential is negative as well as the non-destabilisation of the background when $y_{45}$ and $y_{55}$ sit at maxima are confirmed. 

\begin{figure}[h]
\centering
\includegraphics[width=7cm]{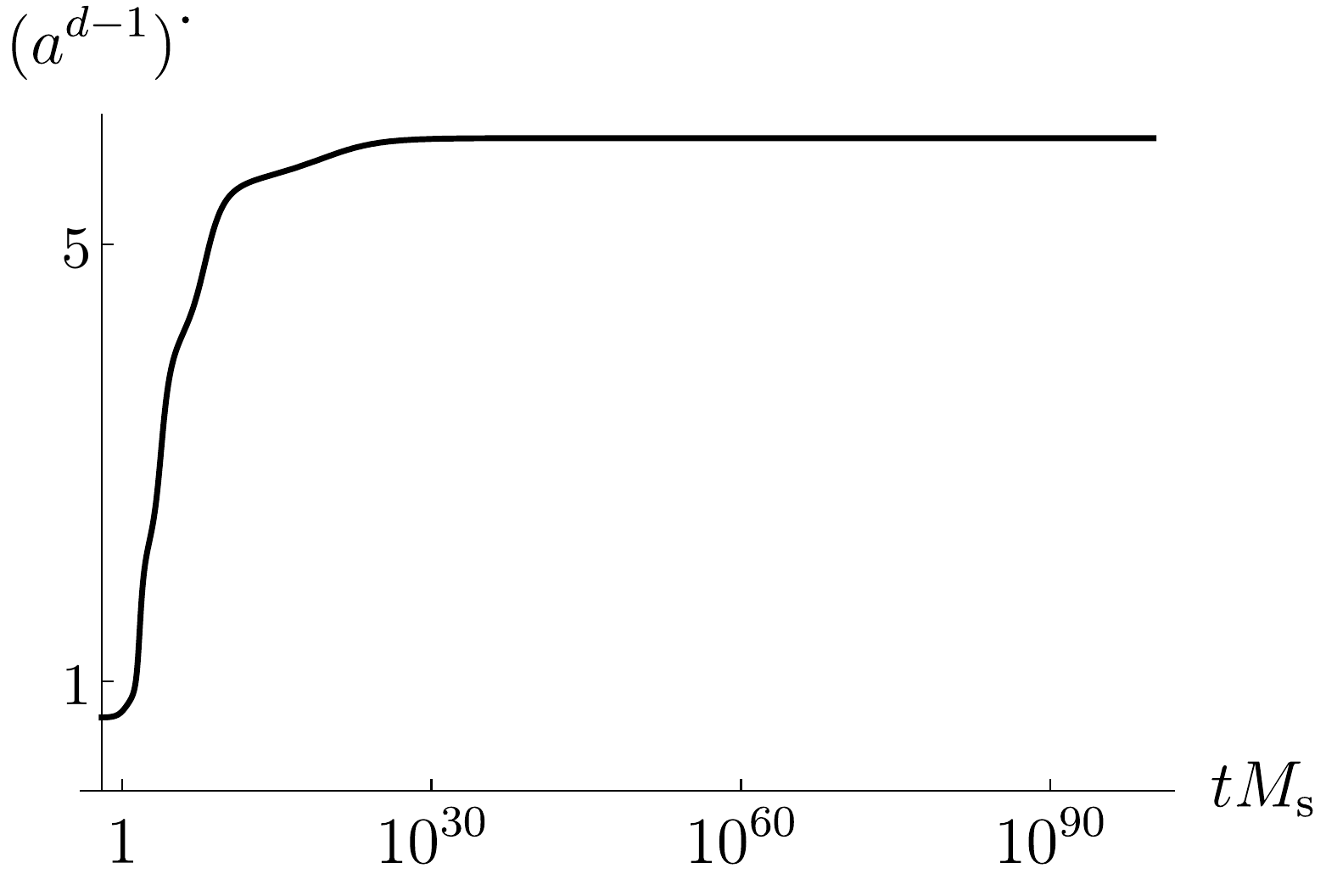} \qquad
\includegraphics[width=7cm]{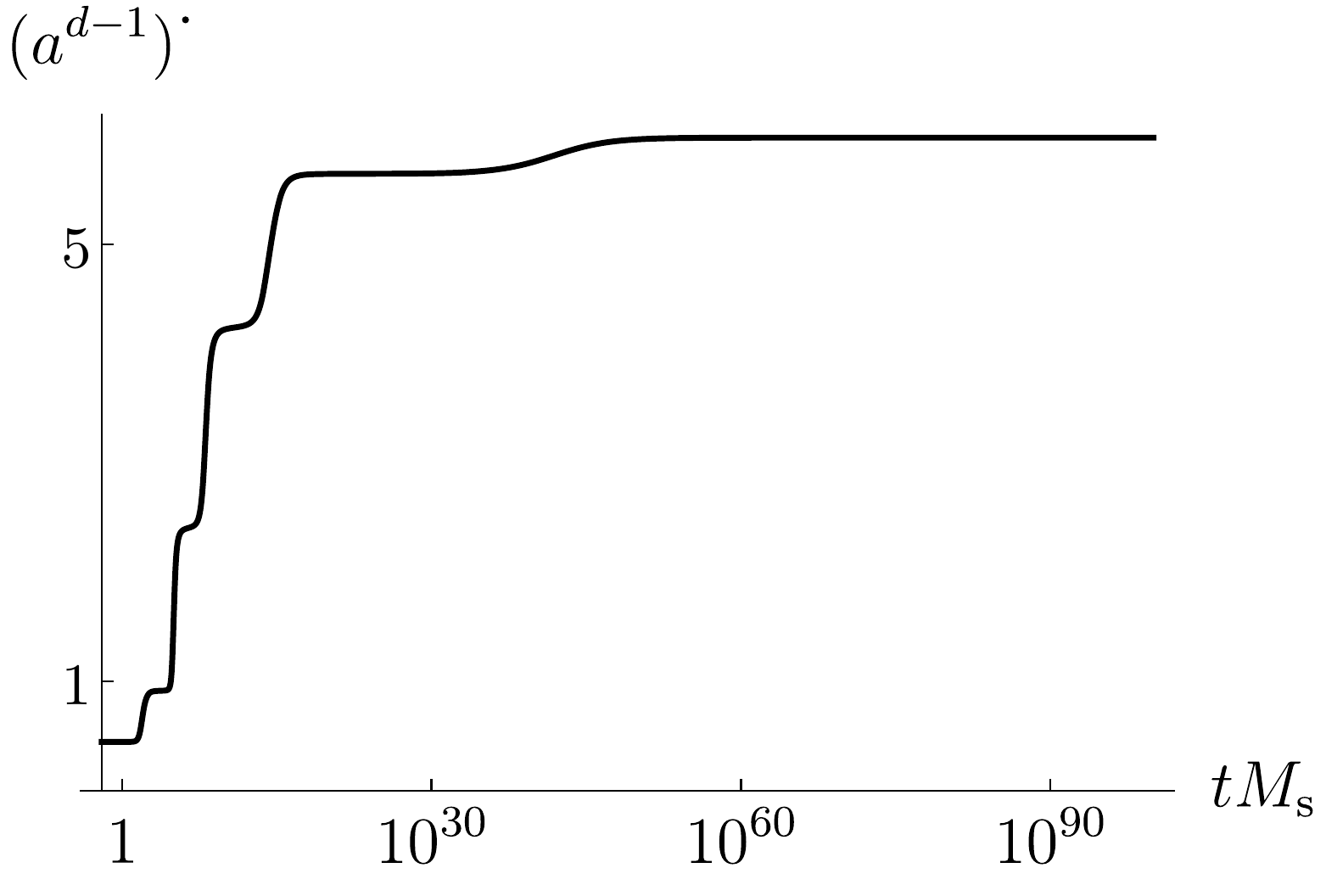}
\begin{picture}(0,0)
\put(-449,99){\mbox{\Large$(i)$}}\put(-226,99){$\mbox{\Large$(ii)$}$}
\end{picture}
\caption{\footnotesize $(i)$~Convergence of $(a^{d-1})^{\displaystyle \cdot}$ towards a constant, signalling the entrance of the universe in QNSR. $(ii)$~When the initial velocities of the Wilson lines are small, the final convergence is reached after successive approximate QNSRs.}
\label{pla}
\end{figure}

On the contrary, if the trajectory of $\Big({c^{\rm sim}_\bot\over \gamma_cc^{\rm sim}_\Phi}, {c_y^{\rm sim}\over \gamma_cc_\Phi}\Big)$ starts outside the shell and/or the initial velocities are high, the scalars $y$ are expected to explore large distances in moduli space. Thus, we are forced to simulate $a(t)$, $\Phi(t)$, $\phi_\bot(t)$, $y_{45}(t)$ and $y_{54}(t)$ using the exact kinetic terms and the full potential~(\ref{po}). For simplicity, we have kept $y_{55}\equiv 0$ in this analysis, which implies $y_{54}$ to be a flat direction. We find that when $\Vone<0$ for some values of $y_{45}$, the initially expanding flat universes stop growing and then collapse, unless the trajectories sit in the tiny shell. As a result, the presence of Wilson lines implies the set of initial conditions that yield ever-expending universes to be drastically reduced, as compared as in the toy model.  On the contrary, when $\Vone\ge 0$ for all $y_{45}$, the point $\Big({c^{\rm sim}_\bot\over \gamma_cc^{\rm sim}_\Phi}, 0\Big)$ is always attracted towards the shell, inside of which it eventually stops. Recalling that in the QNSR we consider the scale factor satisfies $a^{d-1}\sim \#t$, Fig.~\ref{pla}($i$) shows the trajectory and convergence of $(a^{d-1})^{\displaystyle \cdot}$ to a constant. 

Fig.~\ref{pla}($ii$) represents the curve obtained when the initial velocities $\dot y_{45}(0)$ and $\dot y_{54}(0)$ approach zero. A surprising structure of plateaux appears, which can be understood as follows~: 

{$(1)$}  When the initial time-derivatives of the Wilson lines are small, we are essentially back to the toy model, where we have shown that when $\Vone\ge 0$, the universe is always attracted to a QNSR. Thus, $(a^{d-1})^{\displaystyle \cdot}$ converges to its first plateau. 

{$(2)$} However, at this stage, the value of $J_+$ is in most cases negative (we do not a priori sit in the tiny shell). As a result, the kinetic energies of $y_{45}$ and $y_{54}$ end up dominating in Friedmann equation (see the remark below Eq.~(\ref{fri})) and destabilize the above ``approximate'' QNSR.  Since 
\be
{1\over d-1}\, {(a^{d-1})^{\displaystyle \cdot\cdot}\over a^{d-1}}\equiv \dot H+(d-1)H^2={2\over d-2}\, \kappa^2\Vone>0\, ,
\ee
$(a^{d-1})^{\displaystyle \cdot}$ leaves the plateau from above. 

{$(3)$} However, we have shown analytically that the domination of the  kinetic energies of $y_{45}$ and $y_{54}$ cannot last forever (there is no such asymptotic solution). Therefore, these moduli have to release this energy to the rest of the system, and the latter is back to the conditions mentioned in point $(1)$. 

{$(4)$} The process is repeated until $J_+$ becomes positive, \ie the representative point of the system stops in the shell, so that the universe stays in QNSR  for good.  


\section{Conclusion}
\label{cl}

We have shown that the flat cosmological evolutions found at the quantum level in generic no-scale models can be identical, asymptotically, to those found classically. This means that the no-scale structure broken by quantum effects can be restored during the evolution. We have seen on an example that global attractor mechanisms can make the entrance into the QNSR $a\to +\infty$ unavoidable when $\Vone\ge 0$. On the contrary, when the potential can reach negative values,  
the QNSR $a\to +\infty$ is only reached when the initial conditions are chosen in a tiny region of the phase space. 

In QNSR, the universe is dominated by the kinetic energies of the supersymmetry breaking scale $M$, of the dilaton and of the Wilson lines $y_{ij}$, where $i,j$ are not directions involved in the Scherk-Schwarz mechanism. On the contrary, the effective potential $\Vone$ and the kinetic energies of the Wilson lines having indices in Scherk-Schwarz directions are negligible. 

 Finally, the scalars $y_{ij}$ are effectively attracted towards extrema of the effective potential, which are not necessarily minima, while the remaining moduli freeze at arbitrary values. 
 

 \section*{Acknowledgement}
 
I am grateful to  Steve Abel, Carlo Angelantonj, Keith Dienes, Emilian Dudas, Sergio Ferrara, Lucien Heurtier, Alexandros Kahagias and Costas Kounnas  for fruitful discussions. 
My work is partially supported by the Royal Society International Cost Share Award.



\begin{thebibliography}{}
%
%


\bibitem{CFP}
  T.~Coudarchet, C.~Fleming and H.~Partouche,
  ``Quantum no-scale regimes in string theory,''
  Nucl.\ Phys.\ B {\bf 930}, 235 (2018) 
  [arXiv:1711.09122 [hep-th]].
  
  \bibitem{CP}
  T.~Coudarchet and H.~Partouche,
  ``Quantum no-scale regimes and moduli dynamics,''
  Nucl.\ Phys.\ B {\bf 933}, 134 (2018)
  [arXiv:1804.00466 [hep-th]].
  

 
\bibitem{Itoyama:1986ei}
  H.~Itoyama and T.~R.~Taylor,
  ``Supersymmetry restoration in the compactified $O(16) \times O(16)'$ heterotic string theory,''
  Phys.\ Lett.\ B {\bf 186} 129 (1987);\\
  S.~Abel, K.~R.~Dienes and E.~Mavroudi,
  ``Towards a non-supersymmetric string phenomenology,''
  Phys.\ Rev.\ D {\bf 91} 126014 (2015) 
  [arXiv:1502.03087 [hep-th]];\\
S.~Abel, K.~R.~Dienes and E.~Mavroudi,
  ``GUT precursors and entwined SUSY: The phenomenology of stable nonsupersymmetric strings,''
  Phys.\ Rev.\ D {\bf 97}, no. 12, 126017 (2018)
  [arXiv:1712.06894 [hep-ph]];\\
  I.~Florakis and J.~Rizos,
  ``Chiral heterotic strings with positive cosmological constant,''
  Nucl.\ Phys.\ B {\bf 913} 495 (2016) 
  [arXiv:1608.04582 [hep-th]].
  
  
    \bibitem{SNSM}
  C.~Kounnas and H.~Partouche,
  ``Super no-scale models in string theory,''
  Nucl.\ Phys.\ B {\bf 913} 593 (2016)
  [arXiv:1607.01767 [hep-th]];\\
  C.~Kounnas and H.~Partouche,
  ``$\mathcal N=2 \to 0$ super no-scale models and moduli quantum stability,''
  Nucl.\ Phys.\ B {\bf 919} 41 (2017)
  [arXiv:1701.00545 [hep-th]].



    \bibitem{L=0}
  S.~Kachru, J.~Kumar and E.~Silverstein,
  ``Vacuum energy cancellation in a non-supersymmetric string,''
  Phys.\ Rev.\ D {\bf 59} 106004 (1999) 
  [hep-th/9807076]; \\
  G.~Shiu and S.~H.~H.~Tye,
  ``Bose-Fermi degeneracy and duality in non-supersymmetric strings,''
  Nucl.\ Phys.\ B {\bf 542} 45 (1999) 
  [hep-th/9808095];\\
  J.~A.~Harvey,
  ``String duality and non-supersymmetric strings,''
  Phys.\ Rev.\ D {\bf 59} 026002 (1999) 
  [hep-th/9807213];\\
   %
  R.~Blumenhagen and L.~Gorlich,
  ``Orientifolds of non-supersymmetric asymmetric orbifolds,''
  Nucl.\ Phys.\ B {\bf 551} 601(1999) 
  [hep-th/9812158];\\
%
  C.~Angelantonj, I.~Antoniadis and K.~Forger,
  ``Non-supersymmetric type I strings with zero vacuum energy,''
  Nucl.\ Phys.\ B {\bf 555} 116 (1999) 
  [hep-th/9904092];\\
  C.~Angelantonj and M.~Cardella,
  ``Vanishing perturbative vacuum energy in nonsupersymmetric orientifolds,''
  Phys.\ Lett.\ B {\bf 595}  505 (2004)
  [hep-th/0403107];\\
  Y.~Satoh, Y.~Sugawara and T.~Wada,
  ``Non-supersymmetric asymmetric orbifolds with vanishing cosmological constant,''
  JHEP {\bf 1602} 184 (2016) 
  [arXiv:1512.05155 [hep-th]];\\
  Y.~Sugawara and T.~Wada,
  ``More on non-supersymmetric asymmetric orbifolds with vanishing cosmological constant,''
  JHEP {\bf 1608} 028 (2016) 
  [arXiv:1605.07021 [hep-th]];\\  
  %
  S.~Groot Nibbelink, O.~Loukas, A.~M\"utter, E.~Parr and P.~K.~S.~Vaudrevange,
  ``Tension between a vanishing cosmological constant and non-supersymmetric heterotic orbifolds,''
  arXiv:1710.09237 [hep-th].
  
  
  \bibitem{Faraggi}
  A.~E.~Faraggi and M.~Tsulaia,
  ``On the low energy spectra of the nonsupersymmetric heterotic string theories,''
  Eur.\ Phys.\ J.\ C {\bf 54}  495 (2008)
  [arXiv:0706.1649 [hep-th]];\\
  A.~E.~Faraggi and M.~Tsulaia,
  ``Interpolations among NAHE-based supersymmetric and nonsupersymmetric string vacua,''
  Phys.\ Lett.\ B {\bf 683}  314 (2010)
  [arXiv:0911.5125 [hep-th]];\\
  J.~M.~Ashfaque, P.~Athanasopoulos, A.~E.~Faraggi and H.~Sonmez,
  ``Non-tachyonic semi-realistic non-supersymmetric heterotic string vacua,''
  Eur.\ Phys.\ J.\ C {\bf 76}  no.4,  208 (2016)
  [arXiv:1506.03114 [hep-th]].

  
   \bibitem{L2}
  S.~Abel and R.~J.~Stewart,
  ``On exponential suppression of the cosmological constant in non-SUSY strings at two loops and beyond,''
  Phys.\ Rev.\ D {\bf 96} 106013 (2017) 
  [arXiv:1701.06629 [hep-th]];\\
  K.~Aoki, E.~D'Hoker and D.~H.~Phong,
  ``Two-loop superstrings on orbifold compactifications,''
  Nucl.\ Phys.\ B {\bf 688} 3 (2004) 
  [hep-th/0312181];\\
  R.~Iengo and C.~J.~Zhu,
  ``Evidence for nonvanishing cosmological constant in nonSUSY superstring models,''
  JHEP {\bf 0004} 028 (2000) 
  [hep-th/9912074].

  
    \bibitem{SSstring}
R.~Rohm,
``Spontaneous supersymmetry breaking in supersymmetric string theories,''
Nucl.\ Phys.\ B {\bf 237} 553 (1984);\\
C.~Kounnas and M.~Porrati,
``Spontaneous supersymmetry breaking in string theory,''
Nucl.\ Phys.\ B {\bf 310} 355 (1988);\\
  S.~Ferrara, C.~Kounnas and M.~Porrati,
``Superstring solutions with spontaneously broken four-dimensional
  supersymmetry,''
  Nucl.\ Phys.\  B {\bf 304} 500 (1988); \\
S.~Ferrara, C.~Kounnas, M.~Porrati and F.~Zwirner,
``Superstrings with spontaneously broken supersymmetry and their effective theories,''
Nucl.\ Phys.\ B {\bf 318} 75 (1989);\\
  C.~Kounnas and B.~Rostand,
  ``Coordinate-dependent compactifications and discrete symmetries,''
  Nucl.\ Phys.\ B {\bf 341} 641 (1990).
  
    \bibitem{SS}
  J.~Scherk and J.~H.~Schwarz,
  ``Spontaneous breaking of supersymmetry through dimensional reduction,''
  Phys.\ Lett.\  B {\bf 82} 60 (1979). 
    
    
 \bibitem{noscale}
  E.~Cremmer, S.~Ferrara, C.~Kounnas and D.~V.~Nanopoulos,
  ``Naturally vanishing cosmological constant in $\N=1$ supergravity,''
  Phys.\ Lett.\  B {\bf 133} 61 (1983);\\
  J.~R.~Ellis, C.~Kounnas and D.~V.~Nanopoulos,
  ``Phenomenological $SU(1,1)$ supergravity,''
  Nucl.\ Phys.\  B {\bf 241} 406 (1984);\\
  J.~R.~Ellis, A.~B.~Lahanas, D.~V.~Nanopoulos and K.~Tamvakis,
 ``No-scale supersymmetric standard model,''
  Phys.\ Lett.\  B {\bf 134} 429 (1984);\\
  J.~R.~Ellis, C.~Kounnas and D.~V.~Nanopoulos,
  ``No scale supersymmetric GUTs,''
  Nucl.\ Phys.\  B {\bf 247} 373 (1984).


\bibitem{DKS}
  E.~Dudas, N.~Kitazawa and A.~Sagnotti,
  ``On climbing scalars in string theory,''
  Phys.\ Lett.\ B {\bf 694} 80 (2011)
  [arXiv:1009.0874 [hep-th]].
  
  \end{thebibliography}
\end{document}